\documentclass[11pt]{article}
\usepackage{fullpage}
\usepackage{amsmath}
\usepackage{amsfonts}
\usepackage{amssymb}
\usepackage[dvips]{epsfig}

\def\##1{\underline{#1}}
\def\=#1{\underline{\underline{#1}}}

\def\+#1{\underline{\bf #1}}
\def\*#1{\underline{\underline{\bf #1}}}

\def\.{\mbox{ \tiny{$^\bullet$} }}

\def\eps{\epsilon}
\def\epso{\epsilon_{\scriptscriptstyle 0}}
\def\muo{\mu_{\scriptscriptstyle 0}}

\def\c#1{\cite{#1}}
\def\l#1{\label{#1}}
\def\r#1{(\ref{#1})}

\def\le{\left(}
\def\ri{\right)}
\def\les{\left[}
\def\ris{\right]}
\def\lec{\left\{}
\def\ric{\right\}}

\begin{document}

\begin{center}
\Large{\bf {\LARGE  Depolarization regions of nonzero volume
  for anisotropic, cubically nonlinear,   homogenized  nanocomposites }}

\normalsize \vspace{6mm}

Jiajia Cui\footnote{email:  s0457353@sms.ed.ac.uk} and  Tom G.
Mackay\footnote{  email: T.Mackay@ed.ac.uk}

\vspace{4mm}

\noindent{ \emph{School of Mathematics,  University of Edinburgh,\\
 Edinburgh EH9 3JZ,
United Kingdom.} }

\vspace{12mm}

%%% Abstract
{\bf Abstract}\end{center} An implementation of the
strong--permittivity--fluctuation theory (SPFT) is presented in
order to estimate the constitutive parameters of a homogenized
composite material (HCM) which is both cubically nonlinear and
anisotropic. Unlike conventional approaches to homogenization, the
particles which comprise the component material phases are herein
assumed to be small but not vanishingly small. The influence of
particle size on the estimates of the HCM constitutive parameters is
illustrated by means of a representative numerical example. It is
observed that, by taking the nonzero particle size into
consideration, attenuation is predicted and nonlinearity enhancement
is somewhat diminished. In these respects, the effect of particle
size is similar to that of correlation length within the
bilocally--approximated SPFT.

\vspace{8mm}



%\vspace{8mm}

%\newpage

\section{INTRODUCTION}

Formalisms based on the strong--permittivity--fluctuation theory
(SPFT) have been developed to estimate the constitutive parameters
of homogenized composite materials (HCMs), within the realms of both
linear \c{TK81,Genchev,Z94,ML95,MLW00,MLW01} and weakly nonlinear
\c{L01,MLW02a,MLW03,M03} materials. The SPFT approach to
homogenization has the advantage over more conventional approaches,
such as those named after Maxwell Garnett and Bruggeman, that the
distributional statistics of the component material phases is
generally better taken into account \c{Michel00}. For example, the
SPFT is most commonly implemented at the level of the bilocal
approximation, wherein the distributional statistics are described
in terms of a two--point covariance function and its associated
correlation length. Thereby, the bilocally--approximated SPFT
predicts attenuation in the HCM, even when the component phase
materials are nondissipative.

In the  SPFT approach to homogenization, and likewise the Maxwell
Garnett and Bruggeman approaches,  the electromagnetic responses of
the particles which make up the component material phases are
represented by electrically--small  depolarization regions
\c{Michel00}.
 Often, these depolarization regions are taken to be
 vanishingly small \c{M97,WSW99}. The spatial extent of the component
 phase particles is thereby neglected. However, the importance of the
 nonzero spatial extent of the component phase particles
  has been underlined by several  studies
 pertaining to linear isotropic HCMs \c{Doyle,Dungey,SL93a,SL93b,Prinkey,S96}.
 Furthermore,  versions of the SPFT which accommodate  depolarization
 regions of nonzero volume
 were recently developed for
 linear anisotropic \c{M04} and bianisotropic\footnote{In the context of bianisotropic HCMs, the initials SPFT
 stand for strong--property--fluctuation theory.} \c{CM06}
 HCMs.
 While these studies take into
consideration the nonzero spatial extent of the component phase
particles, it is essential that the particle sizes are much smaller
than the electromagnetic wavelengths, in order for the assembly of
component material phases to be regarded as an effectively
homogeneous material \c{Michel00}. Accordingly, at an optical
wavelength of 600 nm, for example, component phase particles with
linear dimensions less than approximately $60$ nm  are envisaged.

In the present study, we extend the consideration of depolarization
regions of nonzero volume into the weakly nonlinear regime for
anisotropic HCMs. Our starting point is  the SPFT for cubically
nonlinear HCMs  which incorporates vanishingly small depolarization
regions \c{M03}. In \S\ref{homog_prelim} the component material
phases and their statistical distributions are described. The linear
and weakly nonlinear contributions to the depolarization dyadics,
arising from depolarization regions of nonzero volume, are presented
in \S\ref{depol_section}. The corresponding bilocally--approximated
SPFT estimate of the HCM permittivity dyadic is given in
\S\ref{homog_section}; and a representative numerical example is
used to illustrate these results in \S\ref{num_studies}. Finally, a
few  closing remarks are provided in \S\ref{conc}.

In the notation adopted, single underlining denotes a 3 vector
whereas double underlining denotes a 3$\times$3 dyadic. The inverse,
determinant, trace and adjoint of  a dyadic $\=M$ are represented as
$\=M^{-1}$, $\mbox{det} \le\, \=M \,\ri$, $\mbox{tr} \le \,\=M
\,\ri$ and $ \=M^{adj} $, respectively. The permittivity and
permeability of free space are written as $\epso$ and $\muo$.

\section{HOMOGENIZATION PRELIMINARIES} \l{homog_prelim}

The homogenization of two
 component material phases, namely phase  $a$ and  phase $b$, is considered. Each
  component
phase is an isotropic dielectric material; and, in general, each is
cubically nonlinear. Thus, the permittivities of the component
material phases are expressed as
\begin{equation} \eps_\ell  = \eps_{\ell 0} +
\chi_\ell \, |\, \#E_{\,\ell} \,|^2, \hspace{30mm} (\ell = a,b),
\l{comp_const}
\end{equation}
with $\eps_{\ell 0}$   being the  linear  permittivity, $\chi_\ell$
the nonlinear susceptibility, and $ \#E_{\,\ell} $ is the electric
field developed inside a region of phase $\ell$ by illumination of
the composite material. The assumption of weak nonlinearity ensures
that $|\,\eps_{\ell 0 }\,| \gg |\, \chi_\ell\,| \, |\, \#E_{\, \ell}
\,|^2$. Notice that nonlinear permittivities of the form
\r{comp_const}  describe electrostrictive materials which can induce
 stimulated Brillouin scattering \c{Boyd}.

 The component material phases $a$ and $b$ are made up of ellipsoidal
 particles. The particles
  all have the same shape and orientation,
  as specified by
 the  shape dyadic
\begin{equation}
\=U = \frac{1}{\sqrt[3]{U_x U_y U_z}} \; \mbox{diag} (U_x, U_y,
U_z),\hspace{25mm} (U_x, U_y, U_z > 0), \l{U}
\end{equation}
which  parameterizes the  particle surface as
\begin{equation} \l{r_e}
\#r^e(\theta,\phi) = \eta \, \=U\.\#{\hat r} \, (\theta, \phi),
\end{equation}
where  $\#{\hat r} \, (\theta, \phi)$ is the radial unit vector
specified by the spherical polar coordinates $\theta$ and $\phi$.
 The size parameter $\eta$ provides a measure of the linear
 dimensions of the ellipsoidal particles.
 It is
assumed that $\eta$ is much smaller than the electromagnetic
wavelengths, but not vanishingly small.

The component phase particles are randomly distributed throughout a
region of volume $V$, which is partitioned into the disjoint regions
of volume $V_a$ and $V_b$ containing phase $a$ and $b$,
respectively. Thus, the component phase distributions are
characterized in terms of statistical moments of the characteristic
functions
\begin{equation}
\Phi_{\ell}(\#r) = \left\{ \begin{array}{ll} 1, & \qquad \#r \in
V_{\ell},\\ & \qquad \qquad \qquad \qquad \qquad \qquad (\ell=a,b) . \\
 0, & \qquad \#r \not\in V_{\ell}, \end{array} \right.
\end{equation}
The first statistical moment of
 $\Phi_{\ell}$
 delivers the volume fraction of phase $\ell$,
  i.e.,  $\langle \, \Phi_{\ell}(\#r) \, \rangle = f_\ell$.
 Plainly,
 $f_a + f_b = 1$. The two--point covariance function
 which constitutes the second statistical moment of $\Phi_{\ell}$
 is taken as the physically--motivated form \c{TKN82}
\begin{equation}
\langle \, \Phi_\ell (\#r) \, \Phi_\ell (\#r')\,\rangle = \left\{
\begin{array}{lll}
\langle \, \Phi_\ell (\#r) \, \rangle \langle \Phi_\ell
(\#r')\,\rangle\,, & & \hspace{10mm}  | \, \=U^{-1}\. \le   \#r - \#r' \ri | > L \,,\\ && \hspace{25mm} \\
\langle \, \Phi_\ell (\#r) \, \rangle \,, && \hspace{10mm}
 | \, \=U^{-1} \. \le  \#r -
\#r' \ri | \leq L\,,
\end{array}
\right.
 \l{cov}
\end{equation}
where $L>0$ is the correlation length. Within the SPFT, the
estimates of HCM constitutive parameters are largely insensitive to
the specific form of the covariance function, as has been shown by
comparative studies \c{MLW02a,MLW01b}.

\section{DEPOLARIZATION DYADIC} \l{depol_section}

Let us focus our attention on a single component phase  particle of
volume $V^e$, characterized by the shape dyadic $\=U$ and size
parameter $\eta$. Suppose that  this particle
 is embedded in
a  comparison medium.
 In consonance with the ellipsoidal
geometry of the component phase particles and the weakly nonlinear
permittivities of the component material phases,  the comparison
medium is a
 weakly nonlinear, anisotropic, dielectric medium characterized by
the permittivity dyadic
\begin{equation}
\=\eps_{\,cm}  = \=\eps_{\,cm0} + \=\chi_{\,cm} \, |\, \#E_{\,HCM}
\,|^2  = \mbox{diag} \le \eps^x_{cm0}, \, \eps^y_{cm0}, \,
\eps^z_{cm0} \ri + \mbox{diag} \le \chi^x_{cm}, \, \chi^y_{cm}, \,
\chi^z_{cm} \ri \, |\, \#E_{\,HCM} \,|^2, \l{cm}
\end{equation}
where $\#E_{\,HCM}$ denotes the spatially--averaged electric field
in the HCM.
 The eigenvectors of
$\=\eps_{\,cm}$ are aligned with those of $\=U$.

 The depolarization dyadic \c{M97}
\begin{equation}
\=D ( \eta ) = \int_{V^e} \, \=G_{\,cm} (\#r) \; d^3 \#r
\l{depol_def}
\end{equation}
provides the electromagnetic response of the ellipsoidal particle
embedded in the comparison medium. Here,  the dyadic Green function
of the comparison medium, namely  $\=G_{\,cm} (\#r)$,  satisfies the
nonhomogeneous vector Helmholtz equation
\begin{equation}
\le \nabla \times \nabla \times \=I
 - \omega^2 \muo \, \=\eps_{\,cm} \ri \. \=G_{\,cm} (\#r -
\#r') =  i \omega \muo \delta \le \#r - \#r' \ri \=I\,. \l{Helm}
\end{equation}
An explicit representation of  $\=G_{\,cm} (\#r)$ is not generally
available \c{W93}, but its  Fourier transform,
\begin{equation}
\={\tilde{G}}_{\,cm} (\#q) = \int_{\#r} \=G_{\,cm} (\#r) \, \exp (-
i \#q \. \#r ) \; d^3 \#r, \l{FT_def}
\end{equation}
may be deduced from \r{Helm}  as
\begin{equation} \l{FTG}
\={\tilde{G}}_{\,cm} (\#q) = - i \omega \muo \le \#q \times \#q
\times \=I  + \omega^2 \muo \, \=\eps_{\,cm} \ri^{-1}.
\end{equation}
By combining  \r{depol_def}, \r{FT_def} and \r{FTG},
\begin{eqnarray}
\=D   &=&
 \frac{\eta}{2 \pi^2} \, \int_{\#q} \frac{1}{q^2} \, \les \frac{\sin
\le q \eta \ri }{q \eta} - \cos \le q \eta \ri \ris \,
\={\tilde{G}}_{\,cm} (\=U^{-1}\.\#q) \; d^3 \#q
\end{eqnarray}
is obtained, after some simplification  \c{M97, MW97}.

\subsection{Depolarization contributions from regions of nonzero  volume}

As in \c{M04}, we express the depolarization dyadic as the sum
\begin{equation}
\=D  = \=D^{\eta = 0} + \=D^{\eta >  0} ,
\end{equation}
where the dyadic
\begin{eqnarray}
 \=D^{\eta =  0} =  \frac{\eta}{2 \pi^2} \, \int_{\#q} \frac{1}{q^2} \,
\les \frac{\sin \le q \eta \ri }{q \eta} - \cos ( q \eta) \ris \les
\lim_{q\rightarrow \infty} \; \underline{\underline{\tilde
G}}_{\,cm} (\=U^{-1}\.\#q) \ris
 \; d^3 \#q
\l{Dinf}
\end{eqnarray}
 represents the depolarization
contribution arising from the region of vanishingly small volume $
\displaystyle{\lim_{\eta \rightarrow 0}} V^e $, whereas  the dyadic
\begin{equation}
 \=D^{\eta > 0 } =  \frac{\eta}{2 \pi^2} \, \int_{\#q}
\frac{1}{q^2} \, \les \frac{\sin \le q \eta \ri }{q \eta} - \cos ( q
\eta) \ris \lec \underline{\underline{\tilde G}}_{\,cm}
(\=U^{-1}\.\#q) - \les \lim_{q\rightarrow \infty} \;
\underline{\underline{\tilde G}}_{\,cm} (\=U^{-1}\.\#q) \ris \ric \;
d^3 \#q  \l{D0}
\end{equation}
 provides the depolarization contribution
arising from the region of nonzero volume $ \le V^e -
 \displaystyle{\lim_{\eta \rightarrow 0}} V^e \ri
 $.

Depolarization dyadics associated with vanishingly small regions
have been  studied  extensively \c{WSW99,MW97}. The volume integral
\r{Dinf} reduces to the $\eta$--independent surface integral \c{M97}
\begin{eqnarray}
\=D^{\eta = 0} &=& \frac{1}{  4 \pi i \omega} \, \int^{2\pi}_{\phi
=0}  \int^\pi_{\theta =0} \les \, \frac{1}{\mbox{tr} \le
\,\=\eps_{\,cm}\.\=A \,\ri} \, \=A \, \ris \;  \sin \theta \;
d\theta\, d\phi , \l{depol}
\end{eqnarray}
 with
\begin{equation}
\=A = \mbox{diag} \,\le \, \frac{\sin^2 \theta \,
\cos^2\phi}{U^2_x},\, \frac{\sin^2 \theta \,
\sin^2\phi}{U^2_y},\,\frac{\cos^2\theta}{U^2_z}\,\ri\,.
\end{equation}
An elliptic function representation for $\=D^{\eta = 0}$  is
available \c{W98} (which simplifies to a hyperbolic
 function representation in the case of a spheroidal depolarization
 region \c{M97}), but for our present purposes the integral representation \r{depol} is more
 convenient.

Depolarization dyadics associated with  small regions of nonzero
volume have lately come under  scrutiny for anisotropic \c{M04} and
bianisotropic \c{CM06} HCMs. As described elsewhere \c{M03,M04},
 by the  calculus of residues \r{D0} reduces to
\begin{eqnarray}
&& \=D^{\eta > 0}= \frac{1}{4 \pi i \omega} \, \=W(\eta), \l{D0_W}
\end{eqnarray}
where the dyadic function
\begin{equation}
  \=W(\eta) = \eta^3  \int^{2\pi}_{\phi
=0}  \int^\pi_{\theta =0} \frac{\sin \theta}{3 \, \Delta} \lec \,
\les \,
 \frac{3 \le \kappa_+ -
\kappa_-  \ri}{2 \eta} + i \le \kappa^{\frac{3}{2}}_+  -
\kappa^{\frac{3}{2}}_-  \ri \ris
  \=\alpha  +
i  \omega^2 \muo \,\le \kappa^{\frac{1}{2}}_+ -
\kappa^{\frac{1}{2}}_-  \ri  \=\beta \, \ric \; d\theta\, d\phi
\l{W_def}
\end{equation}
is introduced. Herein, the dyadics
\begin{eqnarray}
&&  \=\alpha =
  \les 2 \,\=\eps_{\,cm} - \mbox{tr} \le \, \=\eps_{\,cm} \, \ri
\, \=I \, \ris\. \=A - \mbox{tr} \le \,\=\eps_{\,cm}\.\=A\,\ri \,
\=I\,
 -  \, \frac{  \mbox{tr} \le \, \=\eps^{adj}_{\,cm}\.\=A\,\ri -
\les \, \mbox{tr} \le \, \=\eps^{adj}_{\,cm} \, \ri \, \mbox{tr} \le
\, \=A \, \ri \, \ris }{
 \mbox{tr} \le \, \=\eps_{\,cm}\. \=A \, \ri } \,  \=A \,, \nonumber \\
&&
\\
&& \=\beta =
 \=\eps^{adj}_{\,cm} -  \frac{  \det \le \, \=\eps_{\,cm} \, \ri }
{ \mbox{tr} \le \, \=\eps_{\,cm}\. \=A \, \ri} \, \=A \end{eqnarray}
and the scalars
\begin{eqnarray}
&& \Delta = \sqrt{t^2_B - 4 t_A t_C },\\
&& \kappa_\pm = \muo \omega^2 \frac{-t_B \pm \Delta}{2 t_C},
\end{eqnarray}
with
\begin{equation} \left.
\begin{array}{l}
 t_A = \det \le \, \=\eps_{\,cm} \, \ri \vspace{4pt} \\
 t_B =\mbox{tr} \le \, \=\eps^{adj}_{\,cm}\.\=A\,\ri - \les \,
\mbox{tr} \le \, \=\eps^{adj}_{\,cm} \, \ri \, \mbox{tr} \le \, \=A
\, \ri \, \ris \vspace{4pt} \\
 t_C =\mbox{tr} \le \, \=\eps_{\,cm}\. \=A \, \ri \, \mbox{tr} \le
\,  \=A \, \ri \end{array} \right\}.
\end{equation}

Often the approximation
 $ \=D \approx \=D^{\eta = 0}$ is implemented in homogenization studies
 \c{Michel00}. However, studies of isotropic \c{Doyle,Dungey,SL93a,SL93b,Prinkey,S96}, anisotropic \c{M04}
 and bianisotropic \c{CM06} HCMs have emphasized the
importance of
 the nonzero spatial extent of depolarization
regions.

\subsection{Linear and weakly nonlinear depolarization contributions}

We exploit the fact that the comparison medium permittivity \r{cm}
is the sum of a linear part and a weakly nonlinear part  to
similarly express
\begin{eqnarray}
&& \=D = \=D_{\,0} + \=D_{\,1} \, | \, \#E_{\,HCM} \, |^2= \=D^{\eta
= 0}_{\,0} + \=D^{\eta > 0}_{\,0} + \le \=D^{\eta=0}_{\,1} +
\=D^{\eta>0}_{\,1} \ri | \, \#E_{\,HCM} \, |^2,
\end{eqnarray}
where
\begin{equation}
 \=D^{\eta \geqq 0} = \=D^{\eta \geqq 0}_{\,0} + \=D^{\eta \geqq 0}_{\,1} \, | \, \#E_{\,HCM} \,
 |^2.
\end{equation}
The linear and weakly nonlinear contributions to $ \=D^{\eta = 0}$
have been derived earlier \c{M03}; these are
\begin{eqnarray}
&& \=D^{\eta = 0}_{\,0} = \frac{1}{ 4 \pi i \omega  } \,
 \int^{2\pi}_{\phi
=0}  \int^\pi_{\theta =0}    \les \, \frac{1}{\mbox{tr} \le
\,\=\eps_{\,cm0}\.\=A \,\ri} \, \=A \, \ris\,\sin \theta \;  d\theta
\, d\phi ,
 \\&&
 \=D^{\eta = 0}_{\,1} = -
\frac{1}{ 4 \pi i \omega } \,
 \int^{2\pi}_{\phi
=0}  \int^\pi_{\theta =0} \lec \, \frac{\mbox{tr} \le
\,\=\chi_{\,cm}\.\=A \,\ri}{\les \,\mbox{tr} \le
\,\=\eps_{\,cm0}\.\=A \,\ri \,\ris^2} \, \=A \, \ric\,\sin \theta \;
d\theta \, d\phi .
\end{eqnarray}
The linear and weakly nonlinear contributions to $ \=D^{\eta >
0}$~---~and, equivalently, $\=W (\eta ) $~---~follow from
corresponding contributions for an expression analogous to
 \r{W_def} which crops up in the
 bilocally--approximated SPFT \c{MLW00,M03}.
 Thus, we have
 \begin{equation}
\=W (\eta) = \=W_{\,0} (\eta) + \=W_{\,1} (\eta)\, | \, \#E_{\,HCM}
\, |^2 \end{equation}
 with
\begin{equation}
  \=W_{\,0}(\eta) = \eta^3
 \int^{2\pi}_{\phi
=0}  \int^\pi_{\theta =0} \frac{\sin \theta}{3 \, \Delta_0}    \les
\tau_\alpha (\eta)\,
  \=\alpha_{\,0}  + \tau_\beta\, \=\beta_{\,0} \, \ris \;
d\theta \, d\phi \l{W0_def}
\end{equation}
and
\begin{eqnarray}
 \=W_{\,1} (\eta) &=&
 \eta^3
\int^{2\pi}_{\phi =0}  \int^\pi_{\theta =0}
 \frac{\sin \theta}{3 \, \Delta_0}
  \Bigg\{ \tau_{\alpha }(\eta) \le  \=\alpha_{\,1} -
  \frac{\Delta_1}{\Delta_0}
\, \=\alpha_{\,0} \ri
 +
 \tau_{\beta } \le
 \=\beta_{\,1}  - \frac{\Delta_1}{  \Delta_0}
\, \=\beta_{\,0} \ri \nonumber \\ &&
 + \frac{3}{2} \Big[  \le \frac{1}{\eta} + i
\kappa^{\frac{1}{2}}_{0+} \ri \kappa_{1+} - \le \frac{1}{\eta} + i
\kappa^{\frac{1}{2}}_{0-} \ri \kappa_{1-}
 \Big] \=\alpha_{\,0} + \frac{i}{2} \le
\frac{\kappa_{1+}}{\kappa^{\frac{1}{2}}_{0+}} -
\frac{\kappa_{1-}}{\kappa^{\frac{1}{2}}_{0-}} \ri
 \=\beta_{\,0} \Bigg\} \;
d\theta \, d\phi , \nonumber \\ && \l{W1}
\end{eqnarray}
where
\begin{equation}
\left. \begin{array}{l}
 \tau_\alpha (\eta) = \displaystyle{\frac{3 \le \kappa_{0+}
- \kappa_{0-} \ri}{2 \eta} + i \le \kappa^{\frac{3}{2}}_{0+}  -
\kappa^{\frac{3}{2}}_{0-}
\ri} \vspace{8pt}\\
\tau_\beta  = \displaystyle{i  \omega^2 \muo \,\le
\kappa^{\frac{1}{2}}_{0+} - \kappa^{\frac{1}{2}}_{0-}  \ri}
\end{array} \right\}.
\end{equation}
The dyadics $\=\alpha_{\,0}$ and $\=\beta_{\,0}$, and scalars
$\kappa_{0\pm}$ and $\Delta_0$, herein represent the linear parts of
their counterpart dyadics $\=\alpha$ and $\=\beta$, and scalars
$\kappa_\pm$ and $\Delta$, as per \c{M03}
\begin{eqnarray}
&&  \=\alpha_{\,0} =
  \les 2 \,\=\eps_{\,cm0} - \mbox{tr} \le \, \=\eps_{\,cm0} \, \ri
\, \=I \, \ris\. \=A - \mbox{tr} \le \,\=\eps_{\,cm0}\.\=A\,\ri \,
\=I\,
 -  \, \frac{  \mbox{tr} \le \, \=\eps^{adj}_{\,cm0}\.\=A\,\ri -
\les \, \mbox{tr} \le \, \=\eps^{adj}_{\,cm0} \, \ri \, \mbox{tr}
\le \, \=A \, \ri \, \ris }{
 \mbox{tr} \le \, \=\eps_{\,cm0}\. \=A \, \ri } \,  \=A \,, \nonumber \\
&&
\\
&& \=\beta_{\,0} =
 \=\eps^{adj}_{\,cm0} -  \frac{  \det \le \, \=\eps_{\,cm0} \, \ri }
{ \mbox{tr} \le \, \=\eps_{\,cm0}\. \=A \, \ri} \, \=A,\\
&& \kappa_{0\pm} = \muo \omega^2 \frac{-t_{B0} \pm \Delta_0}{2
t_{C0}}, \\&& \Delta_0 = \sqrt{t^2_{B0} - 4 t_{A0} t_{C0} },
\end{eqnarray}
with
\begin{equation}\left. \begin{array}{l}
 t_{A0} = \det \le \, \=\eps_{\,cm0} \, \ri \vspace{4pt} \\
 t_{B0} =\mbox{tr} \le \, \=\eps^{adj}_{\,cm0}\.\=A\,\ri - \les \,
\mbox{tr} \le \, \=\eps^{adj}_{\,cm0} \, \ri \, \mbox{tr} \le \, \=A
\, \ri \, \ris \vspace{4pt} \\
 t_{C0} =\mbox{tr} \le \, \=\eps_{\,cm0}\. \=A \, \ri \, \mbox{tr}
\le \,  \=A \, \ri \end{array} \right\}.
\end{equation}
 Moreover, the weakly nonlinear
contributions to $\=\alpha$, $\=\beta$, $\kappa_\pm$ and $\Delta$
are provided as \c{M03}
\begin{eqnarray}
&& \=\alpha_{\,1} =\les 2 \,\=\chi_{\,cm} -  \frac{t_{B1} t_{C0} -
t_{B0} t_{C1}}{t_{C0} \, \mbox{tr} \le \, \=\eps_{\,cm0}\. \=A \,
\ri }
 - \mbox{tr} \le \,
\=\chi_{\,cm} \, \ri \, \=I \, \ris \. \=A - \mbox{tr} \le
\,\=\chi_{\,cm}\.\=A\,\ri \, \=I  \,, \\ &&
 \=\beta_{\,1} =
 \=\Upsilon - \frac{t_{B1} t_{C0} -
t_{B0} t_{C1}}{t_{C0} \, \mbox{tr} \le \, \=\eps_{\,cm0}\. \=A \,
\ri } \, \=A\,,
\\
&& \kappa_{1 \pm} =  \frac{\omega^2 \le \, -t_{B1} \pm \Delta_1 \,
\ri - 2 t_{C1} \,  \kappa_{0\pm}
}{2 \,  t_{C0}}\,,\\
&&
 \Delta_1 =  \frac{t_{B0} t_{B1} - 2 \, \le \, t_{A1}
t_{C0} + t_{A0} t_{C1} \, \ri}{\Delta_0}\,,
\end{eqnarray}
with
\begin{equation}\left. \begin{array}{l}
 t_{A1} =   \chi^x_{cm} \, \eps^y_{cm0}\, \eps^z_{cm0} +
 \eps^x_{cm0} \, \chi^y_{cm}\, \eps^z_{cm0} +
 \eps^x_{cm0} \, \eps^y_{cm0}\, \chi^z_{cm} \vspace{4pt}\\
 t_{B1} =  \mbox{tr} \le \, \=\Upsilon\.\=A\,\ri - \les \,
\mbox{tr} \le \, \=\Upsilon \, \ri \, \mbox{tr} \le
\, \=A \, \ri \ris \vspace{4pt} \\
 t_{C1} =  \mbox{tr} \le \, \=A \, \ri \,
 \mbox{tr} \le \, \=\chi_{\,cm}\. \=A \, \ri \end{array} \right\},
\end{equation}
and
\begin{equation}
\=\Upsilon = \mbox{diag}\,\le \,
 \chi^y_{cm} \, \eps^z_{cm0} +
\eps^y_{cm0} \, \chi^z_{cm}, \,
 \chi^z_{cm} \, \eps^x_{cm0} +
\eps^z_{cm0} \, \chi^x_{cm}, \,
 \chi^x_{cm} \, \eps^y_{cm0} +
\eps^x_{cm0} \, \chi^y_{cm} \, \ri.
\end{equation}

\section{SPFT ESTIMATE OF HCM PERMITTIVITY } \l{homog_section}

Now that the linear and nonlinear  contributions to the
depolarization dyadic have been established for depolarization
regions of  nonzero volume, we can amalgamate these expressions with
the SPFT for weakly nonlinear anisotropic dielectric HCMs~---~which
is presented elsewhere \c{M03}~---~and thereby estimate the HCM
permittivity.

As a precursor, an estimate of permittivity dyadic of the comparison
medium must first be computed. The Bruggeman homogenization
formalism
 (which is, in fact, equivalent to the lowest--order SPFT \c{MLW00}) is used
 for this purpose. Thus, $\=\eps_{\,cm}$ is found by solving the
 nonlinear equations
\begin{equation} \l{Br_eqn}
 f_a \, \=X_{\,a \mbox{j}} + f_b \, \=X_{\,b \mbox{j}} = \=0\,,
 \hspace{30mm} (\,\mbox{j}=0,1),
 \end{equation}
where
\begin{equation}
\left.
\begin{array}{l}
\=X_{\,\ell \,0} = -i\,\omega\,\le\,\eps_{\ell\, 0}\,\=I -
\=\eps_{\,cm0}\,\ri\.\=\Gamma^{-1}_{\,\ell \,0} \vspace{4pt}
\\
\=X_{\,\ell \,1} = -i\,\omega\, \les \, \le\,g_\ell \,
\chi_{\ell}\,\=I - \=\chi_{\,cm}\,\ri\.\=\Gamma^{-1}_{\,\ell \,0}\,
+ \le\,\eps_{\ell\, 0}\,\=I - \=\eps_{\,cm0}\,\ri\.\=\Lambda_{\,\ell
}\,\ris\,
\end{array}
\ric , \hspace{15mm} (\ell = a,b),
\end{equation}
 are the linear and nonlinear parts, respectively, of the corresponding polarizability
 dyadics.
Herein,
\begin{eqnarray} && \=\Lambda_{\,\ell} = \frac{1}{\det \le \,
\=\Gamma_{\,\ell \, 0}\,\ri}\les \, \mbox{diag} \, \Big( \,
 \Gamma^y_{\ell \, 1} \Gamma^z_{\ell \, 0} + \Gamma^y_{\ell \, 0} \Gamma^z_{\ell \, 1},\,
 \Gamma^z_{\ell \, 1} \Gamma^x_{\ell \, 0} + \Gamma^z_{\ell \, 0} \Gamma^x_{\ell \, 1},\,
 \Gamma^y_{\ell \, 1} \Gamma^x_{\ell \, 0} + \Gamma^y_{\ell \, 0} \Gamma^x_{\ell \, 1}
\,\Big) - \rho_\ell \, \=\Gamma^{-1}_{\,\ell\,0} \, \ris,\nonumber \\
&& \l{e23}
\end{eqnarray}
with
\begin{eqnarray}
\rho_\ell = \Gamma^x_{\ell \, 0} \Gamma^y_{\ell \, 0} \Gamma^z_{\ell
\, 1} + \Gamma^x_{\ell \, 0}
 \Gamma^y_{\ell \, 1}
\Gamma^z_{\ell \, 0} + \Gamma^x_{\ell \, 1} \Gamma^y_{\ell \, 0}
\Gamma^z_{\ell \, 0},
\end{eqnarray}
are expressed in terms of components of the dyadics
\begin{equation}\left. \begin{array}{l}
\=\Gamma_{\,\ell\,0}  =
  \=I + i \omega \, \=D_{\,0}\.\le\,
\eps_{\ell\,0}\,\=I - \=\eps_{\,cm0}\,\ri =
 \mbox{diag} \le \Gamma^x_{\ell\, 0}, \,\Gamma^y_{\ell\,
0},\,\Gamma^z_{\ell \, 0} \ri \vspace{4pt} \\  \=\Gamma_{\,\ell\,1}
=
 i \omega \,\les \, \=D_{\,0}\.\le\,
g_\ell \, \chi_{\ell}\,\=I - \=\chi_{\,cm}\,\ri + \=D_{\,1}\.\le\,
\eps_{\ell\,0}\,\=I - \=\eps_{\,cm0}\,\ri \,\ris  = \mbox{diag}
 \le \Gamma^x_{\ell \, 1}, \,\Gamma^y_{\ell \, 1},\,\Gamma^z_{\ell \, 1} \ri \end{array} \right\}; \l{Gamma1}
\end{equation}
and the local field factor is estimated by \c{LL01}
\begin{equation}
g_\ell =  \left| \,
 \frac{1}{3}\, \les \,
 \mbox{tr} \le \,
\=\Gamma^{-1}_{\,\ell\, 0} \,\ri \ris \,
 \right|^2.
\end{equation}
Estimates of the  $\=\eps_{\,cm0}$ and $\=\chi_{\,cm}$ may be
straightforwardly extracted from \r{Br_eqn} by  recursive schemes;
see \c{M03} for  details.

Finally,
 the bilocally--approximated SPFT estimate of the HCM permittivity dyadic, namely
\begin{equation}
\=\eps_{\,\Omega}  = \=\eps_{\,\Omega 0} + \=\chi_{\,\Omega} \, |\,
\#E_{\,HCM} \,|^2  = \mbox{diag} \le \eps^x_{\Omega 0}, \,
\eps^y_{\Omega 0}, \, \eps^z_{\Omega 0} \ri + \mbox{diag} \le
\chi^x_\Omega, \, \chi^y_\Omega, \, \chi^z_\Omega \ri \, |\,
\#E_{\,HCM} \,|^2, \l{ba}
\end{equation}
is given as \c{MLW00}
\begin{equation} \l{SPFT_Est}
\left. \begin{array}{l} \=\eps_{\,\Omega 0 } = \=\eps_{\,cm0} -
\displaystyle{\frac{1}{i \,\omega}}\,
\=Q^{-1}_{}\.\=\Sigma_{\,0}\vspace{8pt}\\
 \=\chi_{\,\Omega}=
 \=\chi_{\,cm} - \displaystyle{\frac{1}{i \,\omega}}\,\le \,
\=Q^{-1}_{}\.\=\Sigma_{\,1} +\=\Pi\.\=\Sigma_{\,0}\,\ri
\end{array} \right\}.
\end{equation}
Herein, the linear and nonlinear parts of the mass operator are
represented, respectively, by the dyadics
\begin{equation}\left. \begin{array}{ll} \=\Sigma_{\,0} = & \displaystyle{\frac{f_a f_b}{4 \pi i \omega}}\le\,\=X_{\,a0} -
\=X_{\,b0}\,\ri\.\=W_{\,0}(L)\.\le\,\=X_{\,a0} - \=X_{\,b0}\,\ri \vspace{8pt}\\
\=\Sigma_{\,1} = & \displaystyle{\frac{f_a f_b}{4 \pi i \omega}}
\Big[ 2 \le\,\=X_{\,a0} - \=X_{\,b0}\,\ri\.\=W_{\,0} (L)
\.\le\,\=X_{\,a1} - \=X_{\,b1}\,\ri \vspace{4pt} \\ & +
\le\,\=X_{\,a0} - \=X_{\,b0}\,\ri\.\=W_{\,1}(L) \.\le\,\=X_{\,a0} -
\=X_{\,b0}\,\ri \Big]
\end{array} \right\};
\end{equation}
and the dyadic
\begin{equation}
\=\Pi = \frac{1}{\det \le \, \=Q_{\,0}\,\ri}\,\les \, \mbox{diag} \,
\Big( \,
 Q^y_1 Q^z_0 + Q^y_0 Q^z_1,\,
 Q^z_1 Q^x_0 + Q^z_0 Q^x_1,\,
 Q^y_1 Q^x_0 + Q^y_0 Q^x_1
\,\Big) - \nu \, \=Q^{-1}_{\,0} \, \ris\,,
\end{equation}
with
\begin{eqnarray}
\nu = Q^x_0 Q^y_0  Q^z_1 + Q^x_0 Q^y_1 Q^z_0 + Q^x_1 Q^y_0 Q^z_0,
\end{eqnarray}
 is expressed in terms of the components of
\begin{equation}
\left. \begin{array}{l}  \=Q_{\,0} =
 \=I +
\=\Sigma_{\, 0} \. \=D_{\,0}  =
 \mbox{diag} \le  Q^x_0,
\,Q^y_0,\,Q^z_0 \ri \vspace{4pt} \\
 \=Q_{\,1}  = \=\Sigma_{\, 0} \. \=D_{\,1} +
\=\Sigma_{\,1} \. \=D_{\,0} =
 \mbox{diag} \le  Q^x_1,
\,Q^y_1,\,Q^z_1 \ri \end{array} \right\} .
\end{equation}

\section{NUMERICAL STUDIES} \l{num_studies}

The SPFT estimates \r{SPFT_Est} of the HCM linear permittivity and
nonlinear susceptibility are represented by mathematically
complicated expressions. In order to discern the influence of the
size parameter $\eta$,  parametric numerical studies are called for.
To this end, we investigate the following
 representative
example of a homogenization scenario. Let component phase $a$ be a
cubically nonlinear material  with linear permittivity $\eps_{a 0} =
2 \epso$ and nonlinear susceptibility $\chi_a = 9.07571 \times
10^{-12} \epso \, \mbox{m}^2 \mbox{V}^{-2}\; (\equiv 6.5 \times
10^{-4} \; \mbox{esu})$; and component phase $b$ be a linear
material
 with  permittivity $\eps_b
\equiv \eps_{b 0} = 12 \epso$. The eccentricities of the ellipsoidal
component phase particles are specified by $U_x = 1$, $U_y = 3$ and
$U_z = 15$. These choices of parameter values
 facilitate direct comparisons with a previous  investigation
in which the effects of the size parameter $\eta$ were not included
\c{M03}. Results are presented for an angular frequency of $\omega =
\pi \times 10^{15} \mbox{rad s}^{-1}$ (equivalent to a free--space
wavelength of 600 nm).

\begin{figure}[!ht]
\centering \psfull \epsfig{file=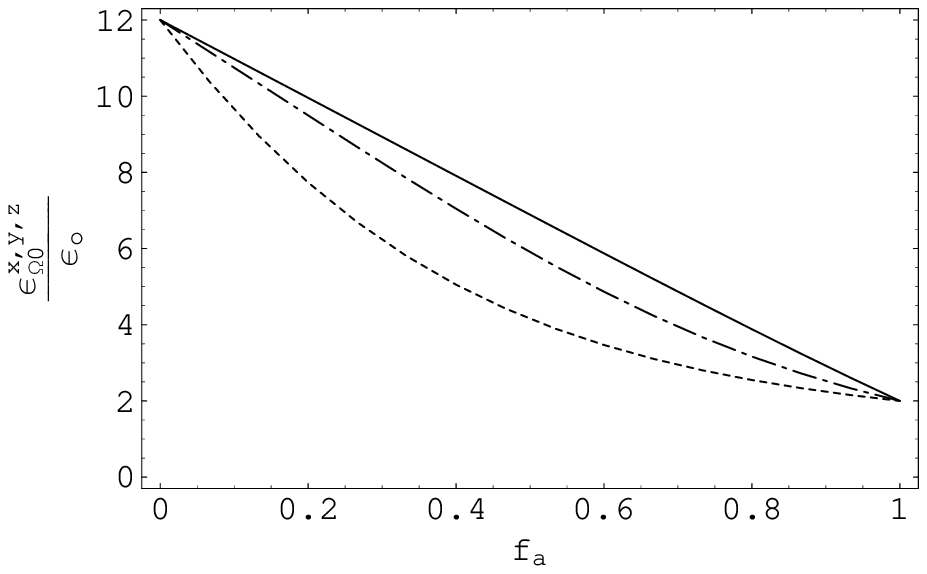,width=2.6in} \hfill
\epsfig{file=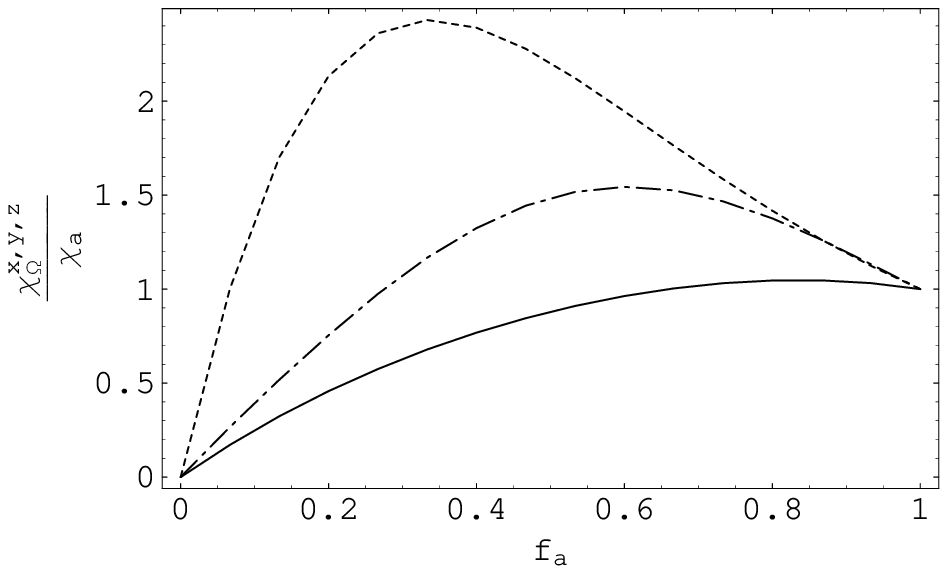,width=2.6in}
 \caption{\label{fig1} The HCM
relative linear permittivity and nonlinear susceptibility parameters
plotted against $f_a$, calculated for $\eta = L = 0$. Key:
$\eps^x_{\Omega 0} /\epso$ and $\chi^x_{\Omega } /\chi_a$ dashed
curves; $\eps^y_{\Omega 0} /\epso$ and $\chi^y_{\Omega } /\chi_a$
broken dashed curves; and $\eps^z_{\Omega 0} /\epso$ and
$\chi^z_{\Omega } /\chi_a$ solid curves. Component phase parameter
values: $\eps_{a 0} = 2 \epso$, $\chi_a = 9.07571 \times 10^{-12}
\epso \, \mbox{m}^2 \mbox{V}^{-2}$, $\eps_b \equiv \eps_{b 0} = 12
\epso$, $U_x = 1$, $U_y = 3$ and $U_z = 15$.
  }
\end{figure}

We begin with the relatively straightforward case where neither the
size parameter nor the correlation length is taken into account;
i.e. $\eta = L = 0$. In this case,  the SPFT estimates of the
constitutive parameters
 are equivalent to those of the conventional Bruggeman formalism
 for weakly nonlinear, anisotropic, dielectric HCMs \c{LL01}.
In Fig.~\ref{fig1}, the HCM linear and nonlinear constitutive
parameters are plotted against volume fraction $f_a$. The HCM linear
permittivity parameters $\eps^{x,y,z}_{\Omega 0}$ uniformly decrease
from   $\eps_{b0}$ at $f_a = 0$ to $\eps_{a0}$ at $f_a = 1$. In
contrast, the HCM nonlinear  susceptibility parameter
$\chi^x_\Omega$, and to  a lesser extent $\chi^y_\Omega$, exceeds
the nonlinear susceptibility of component phase $a$ for a wide range
of values of $f_a$. This \emph{nonlinearity enhancement} phenomenon,
and its potential  for technological exploitation, have been
reported on previously
  for both isotropic \c{L01,MLW03,Boyd96,Liao}  and anisotropic
\c{M03,LL01} HCMs. The anisotropy reflected by the constitutive
parameters, and the nonlinearity enhancement, stems from the
ellipsoidal geometry of the component phase particles.

\begin{figure}[!ht]
\centering \psfull \epsfig{file=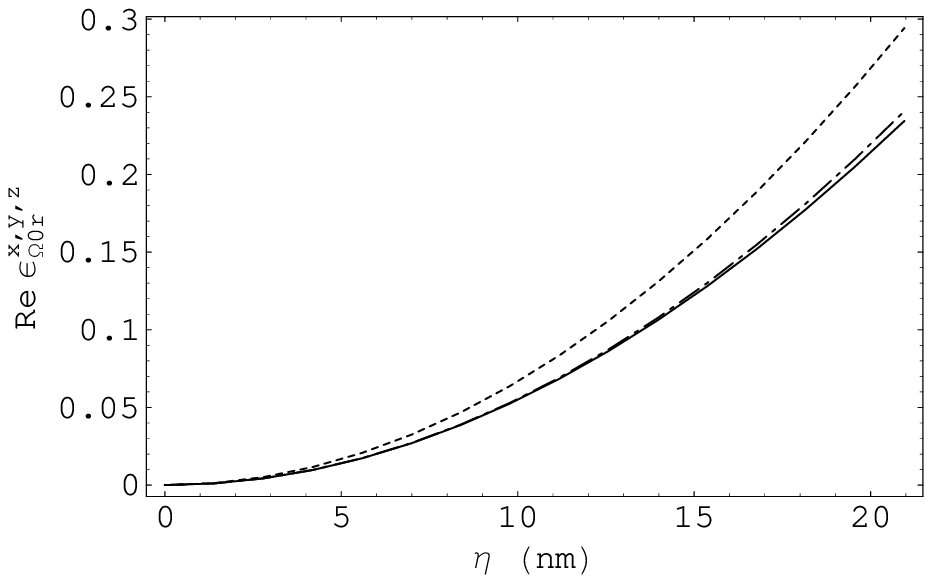,width=2.6in} \hfill
\epsfig{file=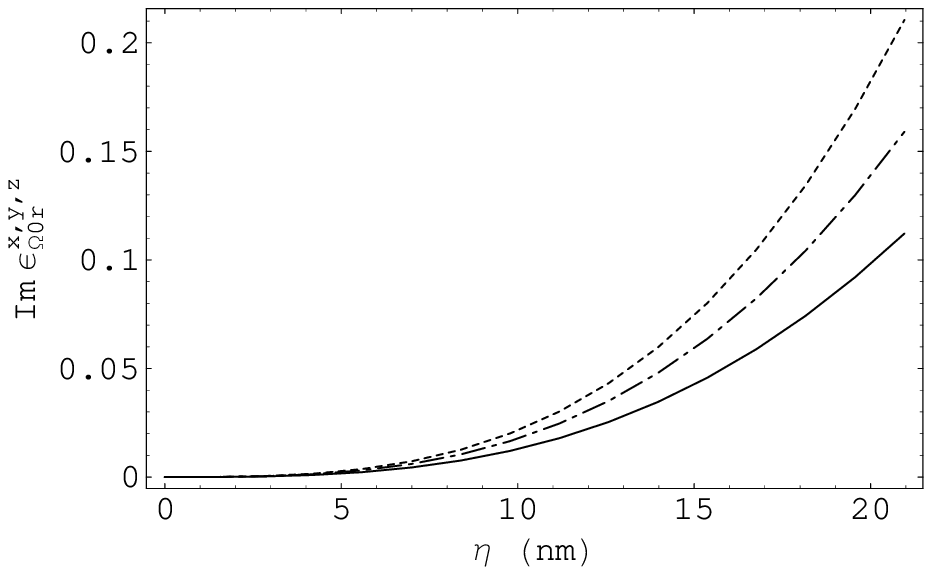,width=2.6in}\\
\epsfig{file=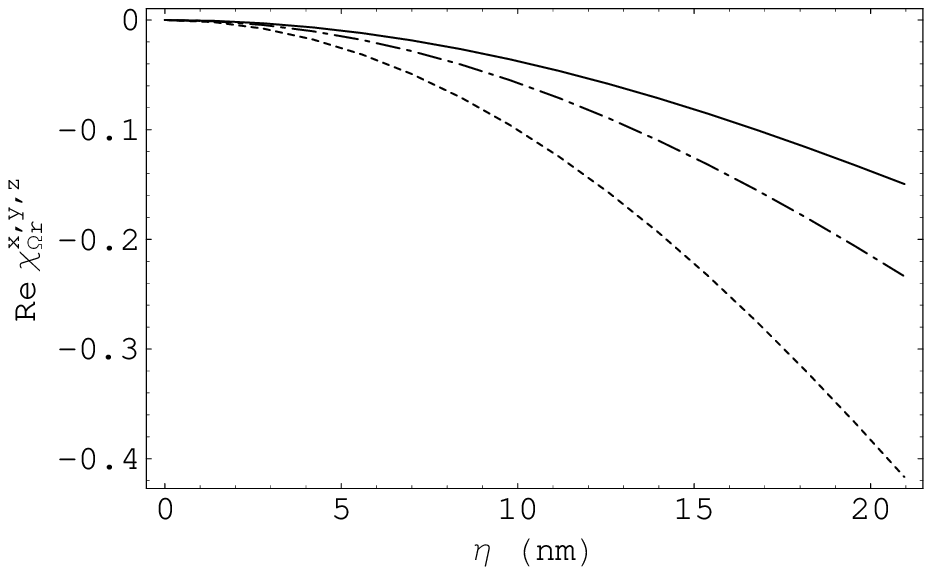,width=2.6in} \hfill
\epsfig{file=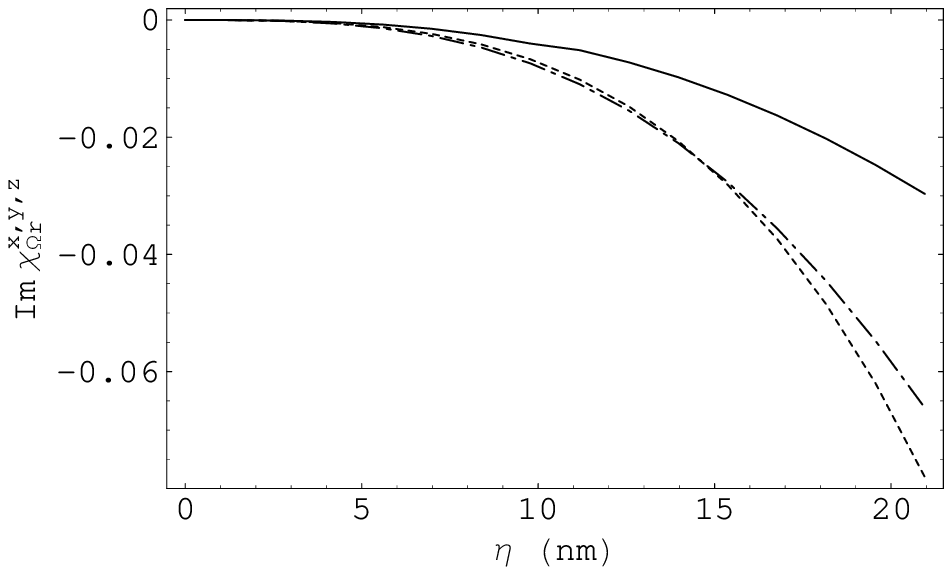,width=2.6in}
 \caption{\label{fig2} Real  and
imaginary parts of the HCM linear  permittivity and nonlinear
susceptibility parameters plotted against $\eta$ (in nm), calculated
for $L=0$ and $f_a = 0.3$. Key: $\eps^x_{\Omega 0r}$ and
$\chi^x_{\Omega r }$ dashed curves; $\eps^y_{\Omega 0r} $ and
$\chi^y_{\Omega r }$ broken dashed curves; and $\eps^z_{\Omega 0r} $
and $\chi^z_{\Omega r} $ solid curves. Component phase parameter
values as in Fig.~\ref{fig1}.
  }
\end{figure}

How does the size parameter $\eta$ influence the estimates of the
HCM constitutive parameters? To answer this question, we fix the
volume fraction at $f_a = 0.3$ and calculate the HCM constitutive
parameters for $ 0 <  \eta < 20$ nm with $L=0$. The presentation of
results is aided by the introduction of the relative constitutive
parameters
\begin{equation}
\left. \begin{array}{l}
 \eps^{n }_{\Omega 0r} = \displaystyle{ \frac{\eps^n_{\Omega
0} - \le \left. \eps^n_{\Omega 0} \right|_{\eta = L= 0} \ri
}{\epso}} \vspace{8pt} \\ \chi^{n }_{\Omega r } = \displaystyle{
\frac{\chi^n_{\Omega } -\le \left. \chi^n_{\Omega }\right|_{\eta =
L= 0} \ri}{\chi_a}} \end{array} \right\}, \hspace{20mm}  (n = x, y,
z),
\end{equation}
which measure the difference between the SPFT estimates calculated
for  $\eta, L \neq 0$ and $\eta = L = 0$. The results
 are plotted in Fig.~\ref{fig2}.
It is notable that the HCM constitutive parameters have nonzero
imaginary parts whereas the component material phases are specified
by real--valued constitutive parameters. As previously described for
linear HCMs \c{M04}, the presence of  nonzero imaginary parts for
the HCM constitutive parameters  may be attributed to radiative
scattering losses associated with the nonzero size of the component
phase particles.
 Plainly,
increasing the size parameter $\eta$ has the effect of increasing
the real and imaginary parts of the HCM linear  permittivity, but
decreasing the real and imaginary parts of the HCM nonlinear
susceptibility. In fact, the influence of the size parameter is very
similar to the influence of the correlation length, as has been
noted for linear HCMs \c{M03}.

\begin{figure}[!ht]
\centering \psfull \epsfig{file=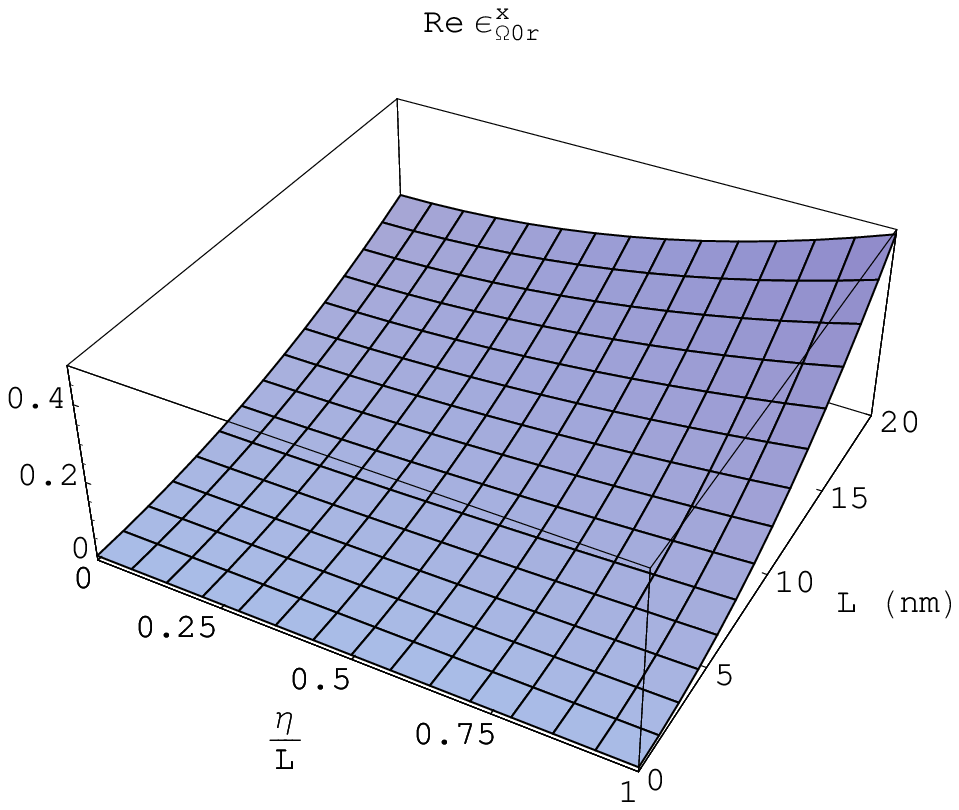,width=2.6in} \hfill
\epsfig{file=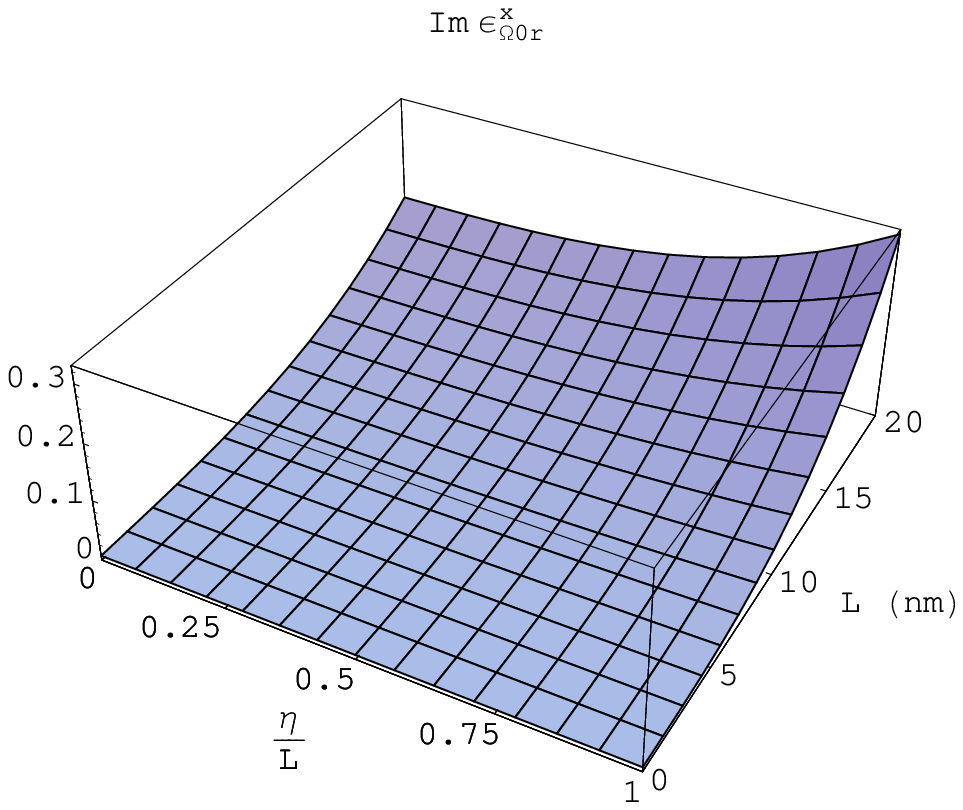,width=2.6in}\\
\epsfig{file=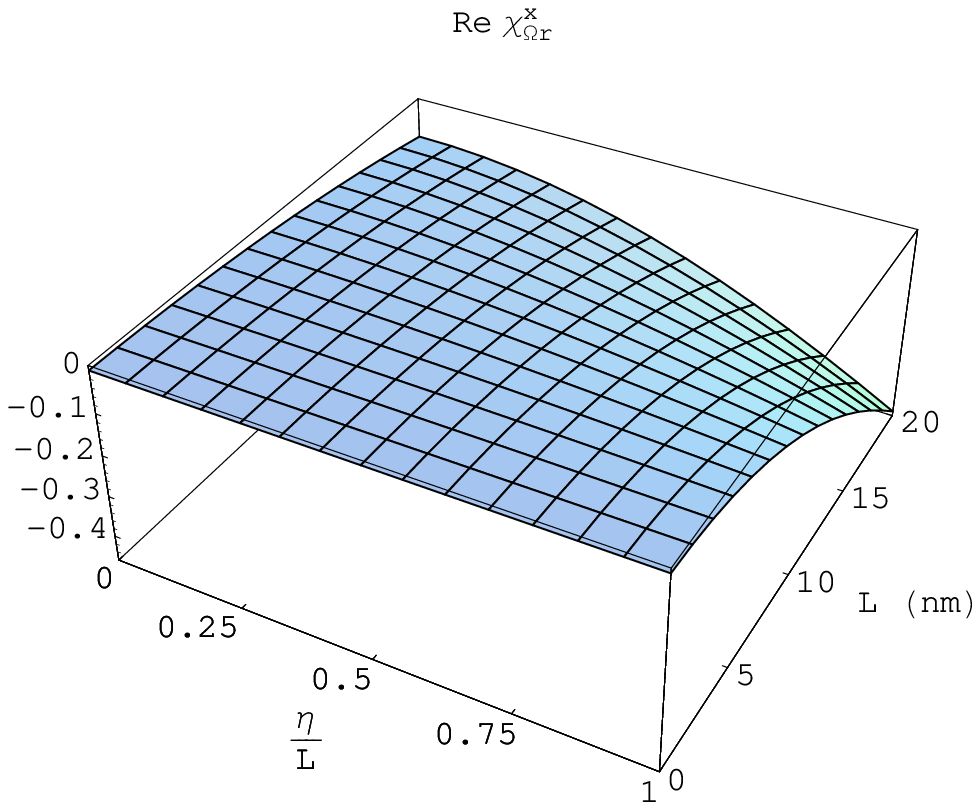,width=2.6in} \hfill
\epsfig{file=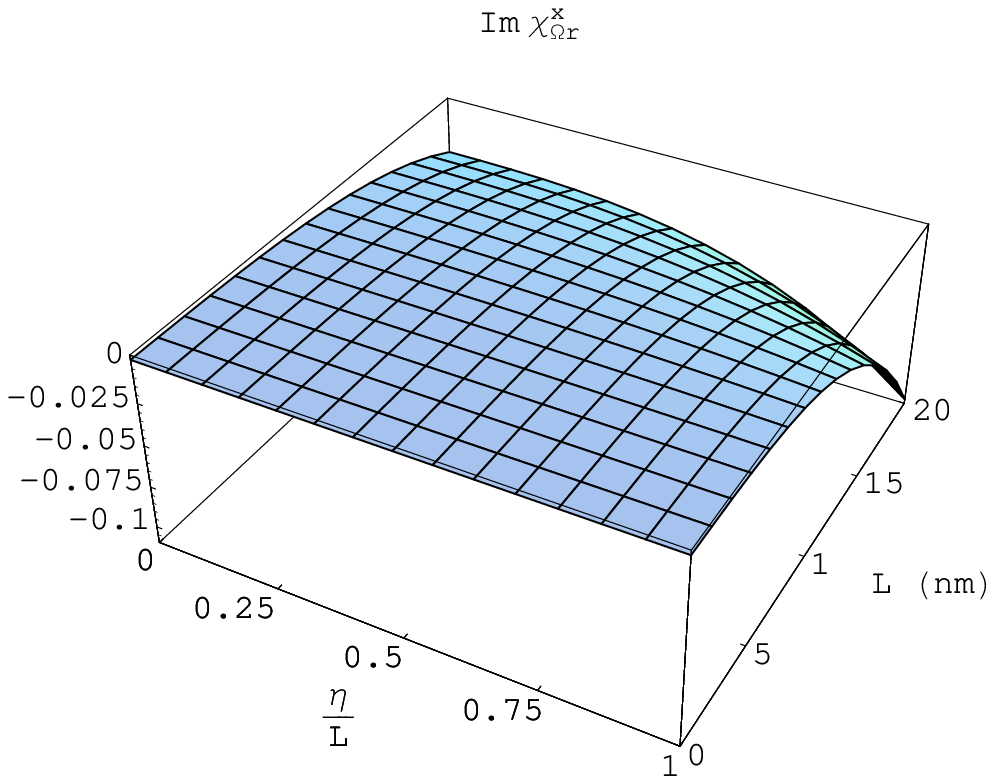,width=2.6in}
 \caption{\label{fig3} Real  and
imaginary parts of the HCM linear  permittivity and nonlinear
susceptibility parameters $\eps^{x}_{\Omega 0r}$ and $\chi^x_{\Omega
r}$ plotted against $ L$ (in nm) and $\eta / L$, calculated for $f_a
= 0.3$.  Component phase parameter values as in Fig.~\ref{fig1}.
  }
\end{figure}

Fig.~\ref{fig2} reveals that by taking
 into consideration the nonzero size of the component
phase particles~---~but not the correlation length~---~the predicted
nonlinearity enhancement is somewhat diminished. We now consider the
estimates of the HCM constitutive parameters when both the size
parameter and the correlation length are  taken into account. In
Fig.~\ref{fig3}, the HCM relative constitutive parameters are
plotted against both $ L$ and $\eta /L$ with the volume fraction
fixed at $f_a = 0.3$. Only the results for $\eps^{x}_{\Omega 0}$ and
$\chi^x_\Omega$ are presented; the corresponding plots for
$\eps^{y,z}_{\Omega 0}$ and $\chi^{y,z}_\Omega$ are  similar. It may
be observed in Fig.~\ref{fig3} that the effects of $\eta$ and $L$
are cumulative insofar as the increase in the real and imaginary
parts
 of  $\eps^{x,y,z}_{\Omega 0}$, and the decrease in the real and imaginary
 parts of $\chi^{x,y,z}_{\Omega}$, which occur as $\eta$ increases,
  become steadily more
  exaggerated as $L  $  increases.

\section{CONCLUDING REMARKS} \l{conc}

The size of the component phase particles can have a significant
bearing upon the estimated constitutive parameters of weakly
nonlinear anisotropic HCMs, within the bilocally--approximated SPFT.
Most obviously, by taking nonzero particle size into consideration,
attenuation is predicted and the degree of nonlinearity enhancement
is somewhat diminished. In respect of both of these effects, the
influence of particle size is similar to the influence of
correlation length. Furthermore, the effects of particle size and
correlation length on both the linear and nonlinear HCM constitutive
parameters are found to be cumulative.

\vspace{10mm} \noindent {\bf Acknowledgements:} JC is supported by a
Scottish Power--EPSRC  Dorothy  Hodgkin Postgraduate Award. TGM is
supported by a \emph{Royal Society of Edinburgh/Scottish Executive
Support Research Fellowship}.


\begin{thebibliography}{}
\bibitem{TK81}
L. Tsang  and J.A.   Kong,
 ``Scattering of electromagnetic waves
from random media with strong permittivity fluctuations," \emph{
Radio Sci.}
 {\bf 16},  303--320 (1981).

\bibitem{Genchev}
Z.D. Genchev,
 ``Anisotropic and gyrotropic version of Polder and
van Santen's mixing formula," \emph{ Waves Random Media} {\bf 2},
99--110 (1992). [doi:10.1088/0959-7174/2/2/001]

\bibitem{Z94}
N.P. Zhuck,
  ``Strong--fluctuation theory for a mean electromagnetic
field in a statistically homogeneous random medium with arbitrary
anisotropy of electrical and statistical properties," \emph{ Phys.
Rev. B} {\bf 50}, 15636--15645 (1994).
[doi:10.1103/PhysRevB.50.15636]

\bibitem{ML95}
B.  Michel  and  A. Lakhtakia,
 ``Strong--property--fluctuation theory
for homogenizing chiral particulate composites," \emph{ Phys. Rev.
E} {\bf 51}, 5701--5707 (1995). [doi:10.1103/PhysRevE.51.5701]


\bibitem{MLW00}
T.G. Mackay, A.  Lakhtakia,  and  W.S. Weiglhofer,
``Strong--property--fluctuation theory for homogenization of
bianisotropic composites: formulation," \emph{ Phys. Rev. E}
 {\bf 62}, 6052--6064 (2000) [doi:10.1103/PhysRevE.62.6052] Erratum  {\bf 63}, 049901
 (2001). [doi:10.1103/PhysRevE.63.049901]

\bibitem{MLW01}
T.G. Mackay,  A.   Lakhtakia,   and  W.S. Weiglhofer,
 ``Third--order
implementation and convergence of the strong--property--fluctuation
theory in electromagnetic homogenisation," \emph{ Phys. Rev. E} {\bf
64}, 066616 (2001). [doi:10.1103/PhysRevE.64.066616]

\bibitem{L01}
A. Lakhtakia,   ``Application of strong  permittivity fluctuation
theory for isotropic, cubically nonlinear, composite mediums,"
\emph{Opt. Commun.} {\bf 192}, 145--151 (2001).
[doi:10.1016/S0030-4018(01)01202-0]

\bibitem{MLW02a}
T.G. Mackay, A. Lakhtakia,  and W.S. Weiglhofer,  ``Homogenisation
of isotropic, cubically nonlinear, composite mediums by the
strong--permittivity--fluctuation theory: third--order
considerations," \emph{Opt. Commun.} {\bf 204}, 219--228 (2002).
[doi:10.1016/S0030-4018(02)01194-X]

\bibitem{MLW03}  T.G. Mackay, A. Lakhtakia, and W.S. Weiglhofer, ``The
strong-property-fluctuation theory for   cubically nonlinear,
isotropic chiral composite mediums,"
 \emph{Electromagnetics} {\bf 23}, 455--479 (2003).
 [doi:10.1080/02726340390203234]

\bibitem{M03}
T.G. Mackay,  ``Geometrically derived anisotropy in cubically
nonlinear dielectric composites,"  \emph{J. Phys. D: Appl. Phys.}
{\bf 36}, 583--591 (2003). [doi:10.1088/0022-3727/36/5/324]

\bibitem{Michel00}
B.  Michel, ``Recent developments in the homogenization of linear
bianisotropic composite materials," in \emph{Electromagnetic Fields
in Unconventional Materials and Structures},  O.N. Singh and A.
Lakhtakia, Eds., pp.39--82, Wiley, New York, NY, USA (2000).

\bibitem{M97}
B. Michel,
 ``A Fourier space approach to the pointwise singularity of
an anisotropic dielectric medium," \emph{ Int. J. Appl. Electromagn.
Mech.} {\bf 8}, 219--227 (1997).

\bibitem{WSW99}
W.S. Weiglhofer, ``Electromagnetic field in the source region: A
review," \emph{Electromagnetics} {\bf 19},  563--578 (1999).

\bibitem{Doyle}
W.T. Doyle,  ``Optical properties of a suspension of metal spheres,"
\emph{ Phys. Rev. B}  {\bf 39}, 9852--9858 (1989).
[doi:10.1103/PhysRevB.39.9852]

\bibitem{Dungey}
C.E. Dungey and  C.F. Bohren,
   ``Light scattering by nonspherical
particles: a refinement to the coupled--dipole method," \emph{
 J. Opt. Soc. Am. A}  {\bf 8}, 81--87 (1991).

\bibitem{SL93a}
B. Shanker and  A. Lakhtakia,
 ``Extended Maxwell Garnett model for
chiral--in--chiral composites," \emph{ J. Phys. D: Appl. Phys.} {\bf
26}, 1746--1758 (1993). [doi:10.1088/0022-3727/26/10/031]

\bibitem{SL93b}
B. Shanker  and  A. Lakhtakia,
 ``Extended Maxwell Garnett formalism for
composite adhesives for microwave-assisted adhesion of polymer
surfaces," \emph{
 J. Composite Mater.} {\bf 27},  1203--1213 (1993).

\bibitem{Prinkey}
M.T.  Prinkey,  A.  Lakhtakia,  and  B. Shanker,
 ``On the extended
Maxwell--Garnett and the extended Bruggeman approaches for
dielectric-in-dielectric composites," \emph{ Optik}  {\bf 96},
25--30 (1994).

\bibitem{S96}
B. Shanker,
  ``The extended Bruggeman approach for chiral--in--chiral
 mixtures," \emph{
J. Phys. D: Appl. Phys.}  {\bf 29},  281--288 (1996).
[doi:10.1088/0022-3727/29/2/002]

\bibitem{M04}
T.G.  Mackay,
 ``Depolarization volume and correlation length in the
homogenization of anisotropic dielectric composites," \emph{ Waves
Random Media } {\bf 14}, 485--498 (2004)
[doi:10.1088/0959-7174/14/4/001] Erratum
 \emph{Waves  Random Complex Media} {\bf 16}, 85 (2006).
 [doi:10.1080/17455030500xxxxxx]

\bibitem{CM06}
J. Cui  and  T.G. Mackay,
 ``Depolarization regions of nonzero volume in bianisotropic homogenized
 composites," \emph{Waves  Random Complex Media} (to appear).
 $\mathsf{http://www.arxiv.org/abs/physics/0608210}$.

\bibitem{Boyd}
R.W. Boyd,  \emph{Nonlinear Optics}, 2nd edition, \S9.3, Academic
Press, London (2003).


\bibitem{TKN82}
L.  Tsang, J.A.  Kong,   and  R.W. Newton,
 ``Application of strong
fluctuation random medium theory to scattering of electromagnetic
waves from a half--space of dielectric mixture,"  \emph{IEEE Trans.
Antennas Propagat.} {\bf 30}, 292--302 (1982).

\bibitem{MLW01b}
T.G. Mackay, A.   Lakhtakia,  and  W.S. Weiglhofer, ``Homogenisation
of similarly oriented, metallic, ellipsoidal inclusions using the
bilocally approximated strong--property--fluctuation theory,"
\emph{Opt. Commun.} {\bf 107}, 89--95 (2001).
[doi:10.1016/S0030-4018(01)01433-X]

\bibitem{W93}
W.S. Weiglhofer,  ``Analytic methods and free--space dyadic Green's
functions," \emph{Radio Sci.} {\bf 28}, 847--857 (1993).

\bibitem{MW97}
B. Michel   and  W.S. Weiglhofer,
  ``Pointwise singularity of dyadic
Green function in a general bianisotropic medium," \emph{ Arch.
Elekron. \"Ubertrag.} {\bf 51}, 219--223 (1997); erratum  {\bf 52},
 31 (1998).

\bibitem{W98}
W.S. Weiglhofer,  ``Electromagnetic depolarization dyadics and
elliptic integrals," \emph{J. Phys. A: Math. Gen.} {\bf 31},
7191--7196 (1998). [doi:10.1088/0305-4470/31/34/019]

\bibitem{LL01}
M.N.  Lakhtakia and A. Lakhtakia, ``Anisotropic composite materials
with intensity--dependent permittivity tensor: the Bruggeman
approach," \emph{Electromagnetics} {\bf 21}, 129--138 (2001).
[doi:10.1080/02726340151134425]

\bibitem{Boyd96}
R.W.  Boyd, R.J.  Gehr,  G.L. Fischer, and J.E. Sipe,  ``Nonlinear
optical properties of nanocomposite materials," \emph{Pure Appl.
Opt.} {\bf 5}, 505--512 (1996). [doi:10.1088/0963-9659/5/5/005

\bibitem{Liao}
H.B.  Liao, R.F.  Xiao, H.  Wang, K.S.  Wong, and G.K.L. Wong,
``Large third--order optical nonlinearity in  $ \mbox{Au:TiO}_2$
 composite films
measured on a femtosecond time scale," \emph{Appl. Phys. Lett.} {\bf
72}, 1817--1819 (1998). [doi:10.1063/1.121193]


\end{thebibliography}
\end{document}